\documentclass[a4paper,12pt]{article}

\usepackage[a4paper]{geometry}
\usepackage[english]{babel}

\usepackage{amsmath}
\usepackage{amsfonts}
\usepackage{amstext}
\usepackage{amssymb}
\usepackage{amsthm}
\usepackage{amscd}
\usepackage{mathrsfs}

\usepackage[pagebackref,draft=false]{hyperref}
\hypersetup{colorlinks,
linkcolor=myrefcolor,
citecolor=mycitecolor,
urlcolor=myurlcolor}

\usepackage[capitalize]{cleveref}
\usepackage{cite}
\usepackage{caption}
\usepackage{etaremune}

\usepackage{xcolor}
\definecolor{myurlcolor}{rgb}{0,0,0.4}
\definecolor{mycitecolor}{rgb}{0,0.5,0}
\definecolor{myrefcolor}{rgb}{0.5,0,0}
\usepackage{graphicx}
\usepackage{tikz}
\usepackage{tikz-cd}
\usetikzlibrary{cd}
\usetikzlibrary{decorations.pathmorphing,shapes}

\usepackage{etoolbox}
\usepackage{enumitem} 
\usepackage[]{latexsym}
\usepackage{braket}
\usepackage{caption}
\usepackage[utf8]{inputenx}
\usepackage[T1]{fontenc}
\usepackage{lmodern}
\usepackage{textcomp}
\usepackage{microtype}
\usepackage{totcount}
\usepackage{blindtext}

\newtheorem*{proof*}{Proof}

\newcommand{\be}{\begin{equation}}
\newcommand{\ee}{\end{equation}}
\newcommand{\bea}{\begin{eqnarray}}
\newcommand{\eea}{\end{eqnarray}}

\numberwithin{equation}{section}
\numberwithin{theorem}{section}

\title{A quantum route to the classical Lagrangian formalism}
\date{}

\author{F. M. Ciaglia$^{1,7}$ \href{https://orcid.org/0000-0002-8987-1181}{\includegraphics[scale=0.7]{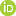}}, F. Di Cosmo$^{2,3,8}$ \href{https://orcid.org/0000-0003-0256-5913}{\includegraphics[scale=0.7]{ORCID.png}}, A. Ibort$^{2,3,9}$ \href{https://orcid.org/0000-0002-0580-5858}{\includegraphics[scale=0.7]{ORCID.png}}, \\ G. Marmo$^{4,5,10}$ \href{https://orcid.org/0000-0003-2662-2193}{\includegraphics[scale=0.7]{ORCID.png}}, L. Schiavone$^{3,4,6,11}$  \href{https://orcid.org/0000-0002-1817-5752}{\includegraphics[scale=0.7]{ORCID.png}}, A. Zampini$^{4,6,12}$ \href{https://orcid.org/0000-0003-0980-6003}{\includegraphics[scale=0.7]{ORCID.png}} \\
\footnotesize{$^{1}$\textit{ Max Planck Institute for Mathematics in the Sciences, Leipzig, Germany}} \\
\footnotesize{$^{2}$\textit{ ICMAT, Instituto de Ciencias Matem\'{a}ticas (CSIC-UAM-UC3M-UCM)}} \\
\footnotesize{$^{3}$\textit{Depto. de Matem\'aticas, Univ. Carlos III de Madrid, Legan\'es, Madrid, Spain}} \\
\footnotesize{$^{4}$\textit{ INFN-Sezione di Napoli, Naples, Italy}} \\
\footnotesize{$^{5}$\textit{ Dipartimento di Fisica ``E. Pancini'', Universit\`a di Napoli Federico II,  Naples, Italy}} \\
\footnotesize{$^{6}$\textit{ Dipartimento di Matematica e Applicazioni "Renato Caccioppoli", Università di Napoli Federico II, Napoli, Italy}} \\
\footnotesize{$^{7}$\textit{ e-mail: \texttt{florio.m.ciaglia[at]gmail.com} and \texttt{ciaglia[at]mis.mpg.de}}} \\
\footnotesize{$^{8}$\textit{ e-mail: \texttt{fcosmo[at]math.uc3m.es}}} \\
\footnotesize{$^{9}$\textit{ e-mail: \texttt{albertoi[at]math.uc3m.es}}} \\
\footnotesize{$^{10}$\textit{ e-mail: \texttt{marmo[at]na.infn.it}}} \\ 
\footnotesize{$^{11}$\textit{ e-mail: \texttt{luca.schiavone[at]unina.it}}}\\
\footnotesize{$^{12}$\textit{ e-mail: \texttt{azampini[at]na.infn.it}}}  
}

\begin{document}

\maketitle

\begin{abstract}
Using the recently developed groupoidal description of Schwinger's picture of Quantum Mechanics, a new approach to Dirac's fundamental question on the role of the Lagrangian in Quantum Mechanics is provided.   It is shown that a function $\ell$ on the groupoid of configurations (or kinematical groupoid) of a quantum system determines a state on the von Neumann algebra of the histories of the system. This function, which we call {\itshape q-Lagrangian}, can be described in terms of a new function $\mathcal{L}$ on the Lie algebroid of the theory. When the kinematical groupoid is the pair groupoid of a smooth manifold $M$, the quadratic expansion of $\mathcal{L}$  will reproduce the standard Lagrangians on $TM$ used to describe the classical dynamics of particles.
\end{abstract}

\section{Introduction: The Lagrangian in Quantum Mechanics}

In this letter a new ``classical'' approximation to the dynamics of a quantum system will be discussed.   It will take advantage of the recently developed groupoid description of quantum mechanical systems emanating from the seminal work by J. Schwinger (see, for instance \cite{Ci19a} - \cite{Ci20b} and references therein).  

There have been many different ways of addressing the relation between the quantum properties of a given system and its  emergent classical description.   In many cases this task takes the form of a ``quantization'' program, that is, guessing the quantum structure of the system from its classical description; in spite of its interest, the only really meaningful question is indeed the converse one: how does the classical structure emerge  from the quantum one?     

At the moment, there is not a clear answer to this question beyond the original proclaims of the founding fathers of Quantum Mechanics.   N. Bohr coined the ``correspondence principle'', that acknowledges the relation between the quantum properties of a system and its classical description  \cite[pages 23-24 and 27-28]{Bo20}: \textit{``The process of radiation cannot be described on the basis of the ordinary theory of electrodynamics, according to which the nature of the radiation emitted by an atom is directly related to the harmonic components occurring in the motion of the system, there is found, nevertheless, to exist a far-reaching correspondence between the various types of possible \underline{transitions} between the stationary states on the one hand and the various harmonic components of the motion on the other hand.... This is equivalent to the statement that, when the quantum numbers are large, the relative probability of a particular \underline{transition} is connected in a simple manner with the amplitude of the corresponding harmonic component in the motion. This peculiar relation suggests a general law for the occurrence of \underline{transitions} between stationary states.'' }

P.A.M. Dirac took a different, and in a sense, much more subtler approach to this question by asking what was the role of the Lagrangian in quantum mechanics \cite{Di33}.   He himself provided a meaningful insight  by relating the Lagrangian to the generating function of canonical transformations of the theory, hence opening the door to subsequent developments brought by R. Feynman and Schwinger, independently.   Each one of them provided different answers to Dirac's conundrum.   Feynman kept the classical Lagrangian $L$ but changed in a drastic way the dynamical description of the theory by providing an explicit expression for the transition amplitudes introducing its path integral approach \cite{Fe48}, while Schwinger opted to change the notion of the Lagrangian function, replacing it with an operator-valued distribution $\mathbf{L}$, and giving  a quantum variational principle description for the dynamics \cite{Sc91}.   In both approaches Bohr's correspondence principle was overridden by the new powerful ideas and relegated to introductory courses on the subject.   The saddle point approximation allows to extract from Feynman's path integral approach the classical trajectories of the system, however the problem persists, as the construction of the path integral uses a classical Lagrangian function $L$.   Where does this Lagrangian come from?

In this note we will propose a new interpretation for the Lagrangian of a quantum theory as a function $\ell$ defined on the groupoid $K$ describing the kinematics of the quantum system.    Such Lagrangian function would allow us to define, on one side, a Feynman-like description of the dynamics of the system by constructing a particular state on a von Neumann algebra of observables of the system and, on the other, a Schwinger-like description of the dynamics of the system by means of a variational principle.   In this Letter, we will focus on the implications of the choice of such a q-Lagrangian function $\ell$ regarding the problem discussed above, that is, showing how a natural ``quasi-classical'' description of the dynamics of the theory emerges using the infinitesimal description of the groupoid $K$ supporting the theory, that is, its Lie algebroid, leaving the dynamical interpretation of the q-Lagrangian $\ell$ to future papers.  

Lie groups are groupoids with a single-element base space, and their Lie algebroids are nothing else than the Lie algebra of the (Lie) group.  Hence, the Lie algebroid description of the dynamics is the natural counterpart of the Lie algebraic description of a dynamics on a group.  
The Lie algebroid of a given Lie groupoid can be integrated by using a natural extension of the exponential map,  and this map allows us to define a Lagrangian function $\mathcal{L}$ on the Lie algebroid starting from a q-Lagrangian $\ell$ on the Lie groupoid. The function $\mathcal{L}$ will be called the c-Lagrangian of the system and it would provide a natural ``classical description'' of the dynamics of the quantum system.    

Lagrangian dynamics on Lie algebroids have been studied in different contexts and their corresponding Euler-Lagrange equations are well understood (see for instance \cite{We96}, \cite{Ma01} and the review \cite{Co06}).    In this Letter it will be shown how the second order approximation to such quasi-classical Lagrangian function allows us to reproduce the well-known Lagrangians found in the description of point particles.   


\section{Quantum systems and groupoids}

On  its most basic terms a quantum system is characterised by the outcomes $x,y,\ldots \in \Omega$ of a family of observables, and by a family of \textit{transitions} $\alpha \colon x \to y$  experienced by the system, whose interpretation is that if the observable $A$ were measured right before the observed transition took place, the outcome would have been $x$, and if measured again right after the transition had taken place, the result would have been $y$.   The outcome $x$ of the transition $\alpha \colon x \to y$ will be called its source and the outcome $y$ its target. The natural axioms satisfied by the family of all possible transitions are those of a groupoid (see \cite{Ci19a} for more details). In particular, transitions compose in a natural way:  the symbol $\beta \circ \alpha$ denotes the transition resulting from the occurrence of the transition $\beta$ right after the first  transition $\alpha$.     Two transitions $\alpha$, $\beta$ can be composed only if the target of the first coincides with the source of the second (note the backwards notation for composition): in this  case they are said to be composable, and such a composition law is associative.   There are unit elements, that is, transitions $1_x\colon x \to x$ such that they do not affect the transition $\alpha \colon x \to y$, when composed on the right, i.e., $\alpha \circ 1_x = \alpha$, or on the left, $1_y \circ \alpha = \alpha$, and whose physical interpretation is that the system remains unchanged during the observation. Finally, the fundamental property that implements Feynmann's principle of microscopic reversibility \cite[page 3]{Fe05}, that is, for any transition $\alpha \colon x \to y$, there is another one, denoted $\alpha^{-1} \colon y \to x$, such that $\alpha^{-1} \circ \alpha = 1_x$ and $\alpha \circ \alpha^{-1} = 1_y$.   The collection of all transitions satisfying the previously enumerated properties is called a (algebraic) groupoid $K$ with space of objects (called in what follows ``outcomes'') $\Omega$.  The map $s \colon K \to \Omega$ assigning to the transition $\alpha \colon x \to y$ the initial outcome $x$ is called the source map, and the map $t \colon K \to \Omega$ assigning to $\alpha$ its final outcome $y$ is called the target map. In the following, we will always consider $K$ and $\Omega$ to be at least locally-compact, Hausdorff topological spaces for which $s,t$ , as well as $x\mapsto 1_{x}$, $\alpha\mapsto\alpha^{-1}$, and $(\beta,\alpha)\mapsto\beta\circ\alpha$, are continuous maps. When $K$ and $\Omega$ are smooth manifolds and all the maps are smooth (in particular, $s,t$ are smooth submersions), we say that the groupoid is a {\itshape Lie groupoid}.

The previous notions provide a natural mathematical setting to Schwinger's `algebra of selective measurements' \cite{Sc91} introduced to provide an abstract setting to the foundations of atomic physics.   Transitions could also be understood in terms of the basic quantum mechanical notion of probability amplitudes because a unitary representation of the given groupoid will associate to them a family of operators directly related to the notion of probability amplitudes or `transition functions' in Schwinger's terminology \cite{Ci19}.   

It is also possible to conceive of a groupoid as an abstraction of a certain experimental setting used to describe the properties of a given system.   For instance, if we consider a charged particle moving on a certain region where detectors have been placed, the triggering of them will correspond to the possible outcomes of the system and the sequence of such triggerings would be the transitions of the system.    Another possible interpretation is offered by earlier descriptions of spectroscopic data.   Actually, as A. Connes suggested \cite{Co94}, the Ritz-Rydberg combination principle of frequencies in spectral lines is rightly the composition law of a groupoid (in this case of a simple groupoid of pairs).   

In what follows, we will look at the groupoid used to describe a certain quantum system as a kinematical object, which means that transitions and outcomes represent just kinematical information obtained from the system without dynamical content, that is, no specific dynamical law is associated to their description.    In this sense, we will say that the groupoid $K$ is a kinematical groupoid. It will also be called the groupoid of ``configurations'' of the system, and it is associated to a specific experimental setting for a (quantum mechanical) system\footnote{Consider, for example, an electron whose motion is analysed through  bubble chambers, or through Stern-Gerlach devices or through  two-slits walls: these settings result in specific (and different from the others) kinematical groupoids.}.

In order to focus  on a specific situation that will be of  particular interest for the purposes of the present letter, consider the groupoid of configurations of a point particle moving in a region $\Omega$.  The outcomes resulting from determining the position of the particle will provide coordinates $x^k$ for the points of the region $\Omega$ with respect to some reference system and the transitions of the system will be  pairs $ \alpha = (y,x)\colon x \to y$ of consecutive detections.    In this case, the natural groupoid that should be used to describe the system will be just the collection of all pairs $(y,x)$, $x,y\in \Omega$, with composition law $(z,y) \circ (y,x) = (z,x)$.   Note that units $1_x$ are diagonal pairs $(x,x)$, and the inverse of the transition $(y,x)$ is $(x,y)$. The resulting groupoid is called the {\itshape pair groupoid} of $\Omega$ and is denoted by $P(\Omega)$.     The region $\Omega$ carries, in general, a natural notion of distance established by the setting prepared by the experimenter.  We may even assume that such distance is given by a metric $ds$ on $\Omega$ (determined, for instance, by the presence of a gravitational field).   Thus, in what follows, we will assume that $(\Omega, \eta)$ is a Riemannian manifold\footnote{We will not deal in this letter with the relativistic descriptions of particles, and address such a problem to a forthcoming paper.} with metric $\eta = \eta_{kl} \, dx^k\,  dx^l$.   

We can even consider a slightly more ambitious setting where, together with the position, the spin of the particle can be measured.  This would imply that a system of orthogonal vectors $e_a(x)$ will be selected at each point $x$, providing an orientation for the Stern-Gerlach-like apparatus used for the task, and each possible transition would amount to a determination of how these vectors are rotated, that is, the transitions of the system will be determined by linear isometries $T_{yx}$ from the tangent  space at the point $x$ to the tangent space at the point $y$ to the manifold $\Omega$, that is $T_{yx}^* \eta_y = \eta_x$, or in local coordinates: $(\eta_y)_{ij} (T_{yx})_k^i (T_{yx})_l^j = (\eta_x)_{kl}$.  In classical terms, we may also think that we are following the possible transitions of a rigid body in the domain $\Omega$. The family of all transitions $T_{yx} \colon x \to y$, constitutes a groupoid $R(\Omega)$ with the natural composition law given by composition of linear maps: $(T_{zy} \circ T_{yx})_a^b = (T_{zy})_a^c (T_{yx})_c^b$.     


\section{Lagrangians and groupoids}

We will take the point of view that the dynamical description of a quantum system is provided by a particular class of states that capture Dirac's insight into the role of the Lagrangian on quantum theories.   The rationale behind this proposal lies in the fact that, given a groupoid $K$ in a large class of groupoids, we may associate a von Neumann algebra $\nu(K)$ to $K$ representing the algebra generated by the (bounded) observables (i.e., self-adjoint elements in $\nu(K)$). This algebra comes  as the closure, in an appropriate  topology, of a suitable convolution algebra of functions on the groupoid, and will be called the (von Neumann) algebra of the groupoid.  In particular, we may follow Connes' theory of integration on groupoids \cite{Co78,Ka82} and define the convolution operation $\star$ on integrable functions on $K$ by setting
$$
(f \star g) (\alpha) = \int_{K^y} f(\gamma) g(\gamma^{-1}\circ \alpha) d \nu^y (\gamma ) \,
$$
where $K^y$ denotes the set of all transitions $\alpha \colon x \to y$, with fixed target $y$, and $\nu^y$ the conditional measure along $K^y$ induced by a measure $\nu$ on $K$. The conjugation operation $f \mapsto f^*$ is defined as $f^*(\alpha) = \overline{f(\alpha^{-1})} \Delta(\alpha^{-1})$ where $\Delta$ is the modular function on $K$, that is $\int f(\alpha) d\nu (\alpha) = \int f(\alpha^{-1}) \Delta (\alpha^{-1}) d\nu (\alpha^{-1})$ for all integrable $f$.  Then, $\nu(K)$ is the weak (or strong) closure of the algebra of integrable functions (closed with respect to the convolution product) represented on the Hilbert space $\mathcal{H}=L^{2}(K,\nu)$ by means of the operators $\left(T_{f}(\Psi)\right)(\alpha)\,:=\,\left(\Delta^{\frac{1}{2}}f\right)\star \Psi(\alpha)$ with $\Psi\in\mathcal{H}$.

By a state of the system described by the groupoid $K$ we mean a state of the von Neumann algebra $\nu(K)$  of observables of the system, that is, a positive normalized linear functional $\rho$ on $\nu (K)$.   A convenient way of describing states is by means of  functions of positive type on the groupoid itself. Suppose that $\varphi$ is a complex-valued function defined on the groupoid $K$ such that
$$
\int_{K} (f^* \star f) (\alpha ) \varphi (\alpha ) d \nu (\alpha) \geq 0 \, ,
$$  
for all integrable functions $f$ on $K$. Then, it can be shown that the functional
$$
\rho_\varphi (f) = \int_{K} f(\alpha) \varphi (\alpha) d\nu (\alpha) \, ,
$$
defines a state on the von Neumann algebra $\nu(K)$ provided that the normalization condition $\int_G \varphi (\alpha) d\nu (\alpha) = 1$ holds (see \cite{Ci21} for more details).

A particularly interesting class of states are those defined by functions of positive type  satisfying a reproducing property (such states were introduced for the first time in \cite{Ci19c}).   It can be shown that if $\varphi$ has the form
$$
\varphi (\alpha ) = \sqrt{p(x) p(y)} e^{ \frac{i}{\hbar} \mathscr{S}(\alpha)} \, ,
$$
for any $\alpha \colon x \to y$, where $p$ is a probability density on $\Omega$, $\hbar$ is a physical constant introduced to make adimensional the argument of the exponential, and the function $\mathscr{S} \colon K \to \mathbb{R}$ satisfies the $\log$-like properties
$$
\mathscr{S}(\alpha \circ \beta ) = \mathscr{S}(\alpha) + \mathscr{S}( \beta ) \, , \qquad \mathscr{S}(\alpha^{-1}) = - \mathscr{S}(\alpha)  \,,  
$$
then $\varphi$ is of positive type and thus determines a state on $\nu (K)$.  We will call such states Dirac-Feynman states, while the function $\mathscr{S}$ will be called an action functional.  

A natural way of constructing an action functional $\mathscr{S}$ consists in considering the groupoid of histories associated to a given groupoid.   In other words, given the groupoid of configurations $K$ of a quantum system, we may consider the space of absolutely continuous paths $w \colon [t_0,t_1] \to K$. Such paths can be composed in a natural way. If $w' \colon [t_1,t_2] \to K$ is another path, then $w' \circ w \colon [t_0,t_2] \to K$  is the path that takes the values $w(t)$ if $t_0 \leq t \leq t_1$  and $w'(t)$ for $t_1 \leq t \leq t_2$, with matching condition $w (t_1) = w'(t_1) $.   Such space of curves can be groupoidified adding the formal inverses of all paths, that is, if $w \colon [t_0,t_1] \to K$ is a path, then $w^{-1}$ is  another path, formally described as a map $w^{-1} \colon [t_1,t_0] \to K$, and such that $w^{-1}(s) =w(s)^{-1}$ (note the opposite orientation of the interval $[t_1,t_0]$ with respect to $[t_0,t_1]$).  The resulting groupoid is called the groupoid of histories $G(K)$ of the configuration groupoid $K$ and its elements are called histories on the groupoid $K$.      Given an integrable function $\ell \colon K \to \mathbb{R}$, we can define the action functional $\mathscr{S}$ on the groupoid of histories $G(K)$  as
$$
\mathscr{S}(w )=  \int_{t_0}^{t_1} \ell (w(s)) ds \, ,
$$
for any history $w$ in $G(K)$.  Notice that the invariance under the ``time-reversal'' operation $\tau\colon \alpha \mapsto \alpha^{-1}$ of the function $\ell$, a property that will be always assumed in what follows, implies that $\mathscr{S}(w^{-1}) = - \mathscr{S}(w)$. 

We can summarise the discussion so far by saying that the choice of a $\tau$-invariant function $\ell$ on the groupoid of configurations $K$ of a quantum system allows to define in a natural way a Dirac-Feynman state on the  groupoid of histories $G(K)$.  We call the function $\ell (\alpha)$ the {\itshape q-Lagrangian} of the theory in accordance with Dirac's terminology of q-numbers and c-numbers.  We must point out at this stage that, even if the q-Lagrangian $\ell$ is an ordinary real-valued function on $K$, it also defines an element of (or is affiliated to) the von Neumann algebra of $K$ when acting on $\mathcal{H}$ by means of the convolution product, hence, it also has a non-ambiguous non-commutative character.   

Let us illustrate the previous ideas with a simple example.   Consider the groupoid of histories of a quantum system described by the groupoid of pairs of the Euclidean 3-space, $E(\mathbb{R}^3) = \{ (\mathbf{y}, \mathbf{x}) \mid \mathbf{x}, \mathbf{y} \in \mathbb{R}^3\}$.  In such case, the groupoid of histories can be identified with the groupoid defined by the family of paths $\gamma \colon [t_0,t_1] \to \mathbb{R}^3$, and a quantum Lagrangian would be given by the function $\ell \colon E(\mathbb{R}^3) \to \mathbb{R}$, $\ell (\mathbf{y}, \mathbf{x}) = \frac{1}{2} || \mathbf{y} - \mathbf{x}||^2$.


\section{The classical  Lagrangian}

Once a q-Lagrangian $\ell$ on the kinematical groupoid $K$ is chosen,   we will see that its ``infinitesimal counterpart'' on the Lie algebroid of $K$ determines a sort of ``classical Lagrangian'', here also called c-Lagrangian, in a sense that will be clear later.\footnote{We are not applying Schwinger's quantum dynamical principle to the q-Lagrangian $\ell$ in order to describe its associated quantum dynamics.   This will certainly constitute the subject of subsequent work.}

As in the case of Lie groups, a Lie groupoid has an infinitesimal description.  In the case of a Lie group $G$, its infinitesimal description is given by its Lie algebra $\mathfrak{g}$, defined either as the space of right (or left) invariant vector fields on $G$ or, equivalently, as the space of tangent vectors at the identity element.   Because of Lie's third theorem the original Lie group $G$ (actually its universal simply connected covering) can be recovered from its Lie algebra $\mathfrak{g}$, and a natural exponential map $\exp \colon \mathfrak{g} \to G$ exists, that agrees with the standard exponential in the case of groups of matrices.     Similar notions hold in the case of Lie groupoids.   If $K \rightrightarrows \Omega$ is a Lie groupoid, we can consider the space of right invariant vector fields on $K$.  Since the right translation $R_\alpha \colon s^{-1}(y) \to s^{-1}(x)$, $\alpha \colon x \to y$, $R_\alpha (\beta) = \beta \circ \alpha$  maps the space of transitions starting at $y$ into the space of transitions starting at $x$, a right-invariant vector field $X$ must be tangent to the fibres of the source map $s$. Moreover, as in the case of Lie groups, such vector fields $X$ turn to be uniquely determined by its values $X_x$ at the units $1_x$, $x \in \Omega$.  Hence we define the Lie algebroid of the Lie groupoid $K$ as the bundle consisting of all $\xi_x$ tangent  to the fibres $K_x$ of the source map $s \colon K \to \Omega$ at the outcomes $x \in \Omega$.   Denoting such collection of tangent vectors as $A(K)$, we have that for each $x\in \Omega$, $\xi_x \in A(K)_x$, if $\xi_x$ is the tangent vector at $x$ of a curve $\gamma (s) \in K$, $\gamma(0) = x$, whose source is $x$ for all $-\epsilon < s < \epsilon$.   

The particular instances of the Riemann groupoid $R(\Omega) $ and the groupid of pairs $P(\Omega)$ provide an excellent illustration of this notion.   In the case of the groupoid of pairs $P(\Omega)$, if we fix $x\in \Omega$, then the set of the transitions whose source is  $x$ are given by all pairs $(y,x)$, $y \in \Omega$.  Hence, the space of tangent vectors at the unit $1_x = (x,x)$ is given by all tangent vectors $\xi_x \in T_x\Omega$, i.e., tangent vectors to curves $(\gamma (s), x)$, $\gamma (0) = x$.    Thus, the Lie algebroid $A(P(\Omega))$ can be identified with the tangent bundle $T\Omega$ over $\Omega$ and the anchor map is the identity map $\mu = \mathrm{id}_\Omega \colon T\Omega \to T\Omega$ (see below).    In the case of the Riemann groupoid $R(\Omega)$, the space of transitions with fixed source $x$  is given by the collection of all linear maps $T_{yx}$ mapping the linear tangent space $T_x \Omega$ isometrically into $T_y\Omega$.   Hence a curve $\gamma (s)$ on $R(\Omega)_x$ can be described as a family of linear maps $T_{x(s)x}$ from $T_x\Omega$ to $T_{x(s)}\Omega$.  For small values of $s$, we can select a smooth frame $\{ \sigma^a(s) \}$, $-\epsilon < s < \epsilon$, in such a way that the maps $T_{x(s)x}$ can be identified with a pair of curves $(x(s), R(s))$ with $x(s)$ a curve in $\Omega$ passing through $x$ and $R(s)$ is a curve of orthogonal matrices. Then, the tangent vector $\xi_x$ to such a curve is just a pair consisting of a tangent vector $v_x$ and a skew-symmetric matrix $S_x$ and the Lie algebroid $A(R(\Omega))$ of the Riemann groupoid is a bundle of Lie algebras over the tangent bundle of the manifold $\Omega$.

The space of all cross-sections $\xi \colon \Omega \to A(K)$, $\xi (x) \in A(K)_x$, of the Lie algebroid $A(K)$, carries a natural Lie algebra structure given by $[\xi , \zeta]_x = [X^\xi, X^\zeta](x)$, where $X^\xi$ denotes the right invariant vector field whose values at $x$ are  $\xi(x)$ (with $x \in \Omega$), while  the bracket on the right hand side of the previous expression is the standard Lie bracket of vector fields.   Finally, the Lie algebroid $A(K)$ carries a natural map (the anchor) $\mu \colon A(K) \to T\Omega$, given by the differential of the target map $t \colon K \to \Omega$ acting upon  vectors $\xi_x$.    The bracket $[\cdot, \cdot ]$ defined before satisfies, in addition to the Jacobi identity, the derivation property
\begin{equation}\label{eq:anchor}
[\xi , f \zeta] = f [\xi , \zeta] + \mu_\xi (f) \zeta \, ,
\end{equation}
where $f$ is an arbitrary smooth function on $\Omega$ and $\mu_\xi := \mu (\xi)$ is a vector field on $\Omega$.   As in the case of Lie algebras, a Lie algebroid $A(K)$ can be characterised by families of structure functions obtained as follows.  Let $\sigma^a$ be a local family of linearly independent cross-sections of $A(K)$ and $x^k$ local coordinates on $\Omega$.  Then, consider the coefficients $C_{ab}^c$ and $\mu_a^k$, defined as
\begin{equation}\label{eq:structure}
[\sigma_a, \sigma_b] = C_{ab}^c \sigma_c \, , \qquad \mu(\sigma_a) = \mu_a^k \frac{\partial }{\partial x^k} \, ,
\end{equation}
that satisfy the non-linear differential equations
$$
\sum_{\mathrm{cyclic\, \, } a,b,c} C_{ad}^e \, C^d_{bc} + \mu_a^k\,  \frac{\partial C_{bc}^e}{\partial x^k} = 0 \, , \qquad 
\mu_a^k\,  \frac{\partial \mu_b^j}{\partial x^k} - \mu_b^k \, \frac{\partial \mu_a^j}{\partial x^k} - C_{ab}^c \, \mu_c^j = 0 \, .
$$
We must point out that all previous notions hold independently whether the bundle $A(K) \to \Omega$ has been constructed starting from a Lie groupoid or not.  In fact, we may define a Lie algebroid $A \to \Omega$ as a vector bundle such that its space of cross-sections carries a Lie algebra bracket $[\cdot, \cdot]$, and there is a map $\mu \colon A \to T\Omega$ satisfying (\ref{eq:anchor}).   
Although an exponential map can be defined for a general Lie algebroid (more precisely, for its Lie algebra of sections), we  concentrate here on a simpler situation where an actual exponential map is constructed for the Lie algebroid $A(K)$ itself.    The natural way to do that is by considering an auxiliary $A$-connection $\nabla$.  As in the standard calculus with covariant derivatives, given a Lie algebroid $A$, an $A$-connection is a bilinear map $\nabla \colon \Gamma (A) \times \Gamma (A) \to \Gamma (A)$ such that $\nabla_{f\xi} \zeta = f \nabla_\xi \zeta$, and $\nabla_\xi (f \zeta) = f \nabla_\xi \zeta + \mu_\xi (f) \zeta$.    If $\sigma_a$ denotes a local basis of cross sections for $A$, the $A$-connection $\nabla$ is characterised by 
 a family of functions $\Gamma_{ab}^c$, defined as $\nabla_{\sigma_a} \sigma_b = \Gamma_{ab}^c \sigma_c$.
As in the case of standard connections, a full covariant calculus can be  developed (see \cite{Cr02} for details).

Given an $A$-connection $\nabla$, we say that the curve $\gamma (s)$ on $A$ such that $\mu (\gamma (s)) = \dot{x}(s)$, with $x(s)$ the projection of the curve on $\Omega$, and $\gamma (0) = x$, is a $\nabla$-geodesic if $\nabla_{\dot{\gamma}} \dot{\gamma} = 0$.  Note that the $\nabla$-geodesic $\gamma(s)$ is determined uniquely by its value $x$ at $s = 0$, and the tangent vector $\xi_x = \dot{\gamma}(0)$. Actually if we write the curve $\gamma (s) = \gamma^a (s) \sigma_a (x)$, then the coefficients $\gamma^a$ satisfy the system of differential equations
$$
\frac{d \gamma^a}{ds} + \Gamma_{bc}^a (x(s))  \gamma^b (s) \gamma^b (s)  = 0 \, , \qquad \frac{d x^k}{ds} = \mu_a^k (x(s)) \gamma^a (s) \, . 
$$

Turning back our attention to the situation where the Lie algebroid $A$ is the Lie algebroid $A(K)$ of a Lie groupoid $K$, fixing an $A(K)$-connection $\nabla$, we define the exponential map $\mathrm{Exp} \colon A(K) \to K$ as $\mathrm{Exp} (x, s\xi_x ) = \gamma_{\xi}(s)$, 
where $\gamma (s)$ is the unique $\nabla$-geodesic such that $\gamma (0) = x$ and $\dot{\gamma}(0) = \xi_x$.    There is a caveat though with respect to the case of the exponential map for Lie groups. Contrarily to the situation with Lie groups, the exponential map $\mathrm{Exp}$ is not defined for all values of the parameter $s$.  Actually, the map $\mathrm{Exp}^\nabla$ is a diffeomorphism on a neighborhood of the space of outcomes (or units) of the theory.     Then, the exponential of the tangent vector $\xi_x$ will be given by $\mathrm{Exp}(x, \xi_x) = \gamma_{\xi_x}(1)$, provided that $\xi_x$ lies in the neighborhood where the geodesic $\gamma_\xi(s)$ is defined for $s=1$ (see \cite{Ma05} for details).    

Given the above analysis, it  makes sense now  to consider the c-Lagrangian associated to the q-Lagrangian $\ell$ as the map $\mathcal{L}$ defined on (an open neighbourhood of the zero section of) the Lie algebroid $A(K)$ by
\begin{equation}\label{eq:classicalL}
\mathcal{L} (x, \xi_x) = \ell \left(\mathrm{Exp} (x, \xi_x/c_K ) \right) \, ,
\end{equation}
and $c_K$ is a constant with the dimensions of $\xi_x$.   In addition, the constant $c_K$ will fix a radius on $A(K)$ such that for all $\xi_x$ whose ``size'' is smaller than $c_K$ the exponential map would be defined. 

If the Lie algebroid $A$ carries a metric along its fibres, say $\eta_x = \eta_{ab}(x) \sigma^a \otimes \sigma^b$, there is a canonical $A$-connection $\nabla^\eta$ associated to it, characterised by the conditions of being torsionless and leaving the metric $\eta$ invariant.    Such $A$-connection coincides with the standard Levi-Civita connection in the case of the groupoid of pairs $G(\Omega)$.   We will assume that the Lie algebroid $A(K)$ of the groupoid $K$ carries a right-invariant metric $\eta$ and the corresponding Levi-Civita $A$-connection $\nabla^\eta$.      

Once a Lagrangian $\mathcal{L}$ is defined on the Lie algebroid $A(K)$ we may describe the dynamics associated to it given by the Euler-Lagrange equations
$$
\frac{d}{dt} \frac{\partial \mathcal{L}}{\partial \xi^a} +   \frac{\partial \mathcal{L}}{\partial \xi^c} \, C_{ab}^c\,  \xi^b - \mu_a^k \, \frac{\partial \mathcal{L}}{\partial x^k}  = 0 \, , \qquad \dot{x}^k = \mu_a^k \, \xi^a \, .
$$
The analysis of this dynamics, introduced by A. Weinstein, has been the subject of intense scrutiny by E. Martinez and others, and has found a wide range of applications (see \cite{Ma01}, \cite{We96} for details).     We leave the discussion of the relation between the dynamics of the c-Lagrangian $\mathcal{L}$ and the original q-Lagrangian $\ell$ to forthcoming papers and focus on the structure of the c-Lagrangian in both general and specific instances.   

Hence, if the kinematical groupoid of the theory $K$ has a natural scale $c_K$ on it, provided that we consider only dynamics with $|| \xi_x || << c_K$, we can approximate the c-Lagrangian $\mathcal{L}$ by truncating its Taylor expansion around the space of outcomes of the theory.  In this sense, we may write the following expansion of $\mathcal{L}$ around the region $\Omega \subset K$
$$
\mathcal{L}(x, \xi_x)  = \mathcal{L}(x, 0_x) + \frac{\partial \mathcal{L} }{\partial \xi^a}(x,0_x) \xi^a + \frac{1}{2} \frac{\partial^2 \mathcal{L} }{\partial \xi^a \xi^b} (x,0_x) \xi^a \xi^b + \mathrm{h.o.t.} 
$$
Taking into account the structure of the function $\mathcal{L}$, Eq. (\ref{eq:classicalL}), we get
\begin{equation}\label{eq:expansion}
\mathcal{L}(x, \xi_x) = \ell (x, x) + \frac{1}{c_K} \frac{\partial_1 \ell }{\partial x^k}(x,x) \mu_a^k  (x) \xi^a + \frac{1}{2c_K^2} \frac{\partial_1^2 \ell} {\partial x^k \partial x^l} (x,x) \mu_a^k \mu_b^l \xi^a \xi^b + O(1/c_K^3) \, ,
\end{equation}
where $ \frac{\partial_1 \ell }{\partial x^k}(x,x)$ denotes the derivative of $\ell$ with respect to the first variable $y$ at the point $(x,x)$.
If we consider the expansion up to second order in (\ref{eq:expansion}), we  get a quadratic function $L$, now defined on the whole Lie algebroid $A(K)$
\begin{equation}\label{eq:L}
L(x, \xi_x) = \frac{1}{2} {\eta}_{ab} (x) \xi^a \xi^b + A_a(x) \xi^a - V(x) \,. 
\end{equation} 
In this expression  $\eta=  \frac{\partial_1^2 \ell} {\partial x^k \partial x^l} (x,x) \, \mu_a^k \, \mu_b^l \, \sigma^a \otimes \sigma^b$ is a quadratic form along the fibres of the Lie algebroid $A(K)$, the linear term reads  $A = A_a\, \sigma^a =  \frac{1}{c} \frac{\partial_1 \ell }{\partial x^k}(x,x) \, \mu_a^k \, \sigma^a$, and $V(x) = - \ell (x,x)$.  We  say that the c-Lagrangian $\mathcal{L}$ (and the q-Lagrangian $\ell$) is regular if the quadratic form $\eta$ above is non-degenerate. In such a case, it defines a metric on the Lie algebroid.   Hence, if the q-Lagrangian $\ell$ is regular, there is a natural metric $\eta_\ell$ associated with it that can be used to construct the exponential map that would allow  to define the second order approximated Lagrangian $L$.   

Clearly, the structure of the second order   approximation   $L$ is reminiscent of the standard form of a Lagrangian describing the motion of a charged point particle in a gravitational field in the presence of an electromagnetic field (see for instance \cite{Ca95} for a thorough discussion on Feynman's inverse problem and the conditions that guarantee that a Lagrangian takes the previous form).  

We  end up the discussion of the classical description of the dynamics of a quantum system by briefly analysing a family of q-Lagrangians which are both natural and physically meaningful.   We  start by considering the groupoid of pairs of a Riemannian manifold $\Omega$ with metric $\eta$, intended to describe  the observed transitions of a particle, an electron for instance, in the region $\Omega$.   There is a natural function on the pair groupoid $P(\Omega) $ which is the two-point function $\ell (y,x)$ given by
$$
\ell (y,x) = \inf_{\gamma \colon (x,s_0) \to (y,s_1)} \frac{1}{2} \int_{s_0}^{s_1} || \dot{\gamma} (s) ||^2 ds \, ,
$$
where $\gamma \colon (x,s_0) \to (y,s_1)$ is any absolutely continuous curve $\gamma \colon [s_0,s_1] \to \Omega$, such that $\gamma (s_0) = x$ and $\gamma (s_1) = y$.   Note that in the particular instance of the $n$-dimensional Euclidean space, then $\ell (y,x) = \frac{1}{2} || y -x ||^2$, as in the example at the end of Sect. 3.   The product of the previous Lagrangian times a constant $mc_K^2$ has  the physical  dimensions of an energy and constitutes the natural candidate for the q-Lagrangian of a free particle moving on the Riemannian manifold $\Omega$.  

In this situation, the Levi-Civita connection $\nabla$ associated to the metric $\eta$ is the natural choice to construct the exponential map $\mathrm{Exp} \colon T\Omega \to \Omega \times \Omega$, that is, $\mathrm{Exp}(x, s\xi_x) =(\exp_x(s\xi_x), x)$, where $\exp$ denotes the standard exponential map (see \cite{La85} for details).     Note that the c-Lagrangian $\mathcal{L}$ (\ref{eq:classicalL}) is a function on (a tubular neighborhood of the zero section of the) tangent bundle $T\Omega$ while its second order approximation (see  Eq. \eqref{eq:L})  is a standard Lagrangian function.  A few computations show that the Lagrangian $\mathcal{L}$ associated to the q-Lagrangian $\ell$  has the form
$$
\mathcal{L}(x, v) = m c_K^2 \ell \left(\mathrm{Exp}(x, v / c_K) \right) = \frac{1}{2} m \,  \eta_{kl}(x) v^k v^l + O(1/c_K^3) \, ,
$$
which coincides with the standard Lagrangian describing the geodetic motion of a particle on $\Omega$.


\section{Conclusions and discussion}

A new approach to analyse the question raised by Dirac on the role of the Lagrangian function in Quantum Mechanics  is presented; it    is based on the groupoidal formulation of Schwinger's algebraic description of Quantum Mechanics.   It has been shown that a choice of a function $\ell$ on the kinematical groupoid describing a quantum system, called a {\itshape q-Lagrangian}, emerges from a state in a particularly interesting class defined by functions of positive type satisfying a reproducing property. This q-Lagrangian becomes a natural candidate to determine the dynamics of the corresponding quantum system according to Schwinger's variational principle.    The function $\ell$, even if it is an ordinary real-valued function on the groupoid of the system,  defines an (affiliated) element of the von Neumann algebra of the groupoid, showing in this way its true non-commutative origin in agreement with Schwinger's notion of quantum Lagrangian.

If the kinematical groupoid is a Lie groupoid, one may exploit the associated Lie algebroid to provide an infinitesimal description of the q-Lagrangian that brings in a  classical-like flavour.   In the particular instance that the q-Lagrangian $\ell$ is regular, there is a canonical exponential map that allows to translate the q-Lagrangian function $\ell$ into a c-Lagrangian $\mathcal{L}$ defined on (a tubular neigborhood of the space of outcomes of) the Lie algebroid of the theory. Hence a classical-like  dynamics is provided by the corresponding Euler-Lagrange equations.   Such dynamical behaviour would only be defined, in principle, for  ``small velocities'' and would only account for a restricted description of the full quantum dynamics.     It is shown that the quadratic approximation to the c-Lagrangian $\mathcal{L}$ produces a function on the Lie algebroid whose explicit form is reminiscent of the standard Lagrangian for particles moving on a gravitational background under an electromagnetic field.

In the particular instance of the groupoid of pairs of a Riemann manifold that would provide the natural setting to describe the free motion of a particle on a curved background, the q-Lagrangian $\ell$ is defined to be the natural two-point function on the Riemannian manifold, and the fourth-order approximation of the associated c-Lagrangian is the standard Lagrangian describing the geodetical motion of a classical particle on $\Omega$. Further examples of interest, like the Riemann groupoid suitable to take into account the spin of particles and the corresponding groupoid for a space time $\mathscr{M}$ with Lorentzian metric $\eta$, will be discussed elsewhere.  

The analysis of the quantum dynamics determined by a q-Lagrangian $\ell$ will be the subject of forthcoming articles and the relation with Lagrangian dynamics on Lie algebroids will be discussed in detail.  We also hope  to show that this setting is the appropriate one to address Dirac query that starts from the observation that each solution of Hamilton-Jacobi equation (corresponding to one state of motion in the quantum theory) gives rise to a family of solutions of Hamilton's equations \cite{Di51}, so presumably the family has some deep significance in nature, not yet properly understood. In particular, the problem of reconstructing the quantum dynamics from the classical one will be analysed from the perspective of the integration of the Lie algebroid of a given groupoid \cite{Cr03}.  This program can be considered as a natural ``quantization'' program in the groupoidal setting for Quantum Mechanics.


\section*{Acknowledgements}
We acknowledge the  support provided by the MINECO research project MTM2017-84098-P and QUITEMAD++, S2018/TCS-A4342, the financial support from the Spanish Ministry of Economy and Competitiveness through the Severo Ochoa Programme for Centres of Excellence in RD(SEV-2015/0554), and the  support provided by the Santander/UC3M Excellence Chair Programme 2019/2020. It is a pleasure for us to thank the Istituto Nazionale di Fisica Nucleare (INFN) and the Gruppo Nazionale di Fisica Matematica (INDAM), Italy. L.S. would like to thank the support provided by Italian MIUR through the Ph.D. Fellowship at Dipartimento di Matematica R.Caccioppoli.


\begin{thebibliography}{999}
%
\bibitem{Bo20}  N. Bohr.  \textit{On the series spectra of the elements}.  Lecture before the German Physical Society in Berlin (27 April 1920). Also found in Niels Bohr Collected Works, Vol. 3: The Correspondence Principle (1918--1923), J. R. Nielsen (ed.). Amsterdam: North-Holland Publishing.

\bibitem{Ca95} J.F. Cari\~nena, A. Ibort, G. Marmo, A. Stern. \textit{The Feynman problem and the inverse problem for Poisson dynamics}. Physics Reports, \textbf{263}(3)  (1995) 153--212.

\bibitem{Ci19} F.M. Ciaglia, A. Ibort, G. Marmo. \textit{A gentle introduction to Schwinger's picture of Quantum Mechanics}. Mod. Phys. Let. A. \textbf{33}(20), 1850122  (2018).   

\bibitem{Ci19a}  F.M. Ciaglia, A. Ibort, G. Marmo. \textit{Schwinger's Picture of Quantum Mechanics I: Groupoids}. Int. J. Geom. Met. Mod. Phys., \textbf{16}, 1950119  (2019).

\bibitem{Ci19b}  F.M. Ciaglia, A. Ibort, G. Marmo. \textit{Schwinger's Picture of Quantum Mechanics II: Algebras and observables}.  Int. J. Geom. Met. Mod. Phys., \textbf{16} (9), 1950136 (2019).

\bibitem{Ci19c}  F.M. Ciaglia, A. Ibort, G. Marmo. \textit{Schwinger's Picture of Quantum Mechanics III: The statistical interpretation}. Int. J. Geom. Met. Mod. Phys., \textbf{16}(11) 1950165 (2019).

\bibitem{Ci20a}  F.M. Ciaglia, F. Di Cosmo, A. Ibort, G. Marmo. \textit{Schwinger's Picture of Quantum Mechanics IV: Composite systems}. Int. J. Geom. Met. Mod. Phys. \textbf{17} (4), 2050058  (2020).

\bibitem{Ci20b}  F.M. Ciaglia, F. Di Cosmo, A. Ibort, G. Marmo. \textit{Schwinger's Picture of Quantum Mechanics}. Int. J. Geom. Met. Mod. Phys. \textbf{17} (4), 2050054  (2020).

\bibitem{Ci21} F. Ciaglia, F. Di Cosmo, A. Ibort, G. Marmo, L. Schiavone, A. Zampini. \textit{On the von Neumann algebra of groupoids and the type of quantum systems}. In preparation.

\bibitem{Co78} A. Connes. \textit{Sur la theorie noncommutative de l'integration}. In Lecture Notes in Mathematics, vol. \textbf{725}, Berlin, Springer (1978).

\bibitem{Co94} A.~Connes, \textit{Noncommutative geometry}, Academic Press (1994).

\bibitem{Co06} J. Cort\'es, M. De Leon, J.C. Marrero, D.M. De Diego, E. Martinez. \textit{A survey of Lagrangian mechanics and control on Lie algebroids and groupoids}. Int. J.  Geom. Met. Mod. Phys., \textbf{3}(03), (20026) 509--558.

\bibitem{Cr02} M. Crainic, R. Loja Fernandes. \textit{Lie algebroids, Holonomy and Characteristic Classes}.  Adv. in Maths., \textbf{170}, (2002) 119--179.

\bibitem{Cr03} M. Crainic, R. Loja Fernandes.  \textit{Integrability of Lie brackets}.  Ann. of Maths., \textbf{157} (2003) 575--620.
%
\bibitem{Di33} P.A.M. Dirac.  \textit{The Lagrangian in Quantum Mechanics}
Physikalische Zeitschrift der Sovietunion, Band 3, Heft 1 (1933). 

\bibitem{Di51} P.A.M. Dirac. Canadian Journal of Mathematics, \textbf{3} (1951) 1-- 23. 

\bibitem{Fe48} R. P. Feynman. \textit{Space-time approach to non-relativistic quantum mechanics}.  Rev. Mod. Phys., \textbf{20} (1948) 367--387.

\bibitem{Fe05} R. P. Feynman.  \textit{Feynman's Thesis: A New Approach to Quantum Theory}.   Editor L. M. Brown, World Scientific (2005).  Reprinted from R. P. Feynman, \textit{The principle of least action in Quantum Mechanics} (1942). 

\bibitem{Ka82} D. Kastler.  \textit{On A. Connes Noncommutative Integration Theory}. Commun. Math. Phys., \textbf{85}, (1982) 99-120.

\bibitem{La85} S. Lang. {\it Differentiable manifolds}.  Springer-Verlag, New York (1985).

\bibitem{Ma05} K.  Mackenzie. \textit{General Theory of Lie groupoids and Lie algebroids}.  London Mathematical Society Lecture Note Series \textbf{213}, Cambridge Univ. Press (2005).

\bibitem{Ma01} E. Martinez. \textit{Lagrangian Mechanics on Lie algebroids}, Acta Appl. Math., \textbf{67} (2001), 295--320.

\bibitem{Sc91}  J. Schwinger.  \textit{Quantum Kinematics and Dynamics}.  Advanced Book Classics, Frontiers in Physics Series. Perseus Books Group, New York (1991).

\bibitem{We96} A. Weinstein. \textit{Lagrangian Mechanics and groupoids}, Fields Inst. Comm. \textbf{7} (1996), 207--231.

\end{thebibliography}
\end{document}